\documentstyle[twoside,fleqn,espcrc2,epsfig]{article}
\def\etal    {\it et al.\rm}

\newcommand{\AmS}{{\protect\the\textfont2
  A\kern-.1667em\lower.5ex\hbox{M}\kern-.125emS}}

\title{Tau2000 Conference Summary}

\author{K.K. GAN\address[]{Department of Physics,
        The Ohio State University,
        Columbus, OH 43210,
        U.S.A.}}
       
\begin{document}

\begin{abstract}
This is a review of some of the highlights of the 6th workshop on the
physics of the $\tau$ lepton and its neutrino.
This includes the test of lepton universality, measurement of Lorentz Structure,
study of hadronic decays, direct evidence of tau neutrino,
status of neutrino oscillations, and search for neutrinoless decays.
This review concludes with a look at the prospect for $\tau$ physics
in the future.
\vspace{1pc}
\end{abstract}

\maketitle

\section{Introduction}

The $\tau$ lepton provides an unique laboratory to test the Standard Model.
The lepton can be studied as the decay product of the $W$ and $Z$ bosons.
Comparison of the decay rates to those for $e$ and $\mu$ allows a test
of lepton universality.
Alternatively, we can examine the decay products of the $\tau$ lepton.
Its large mass allows decays into many channels, providing many
venues to challenge the Standard Model.
This includes the test of lepton universality and measurement of
Lorentz Structure in the leptonic decays, tests of QCD, CVC and isospin
symmetry in the hadronic decays, and the search for neutrinoless decays.
In this paper, I will review the tests of lepton universality, measurement of
Lorentz Structure, study of hadronic decays, and search for neutrinoless decays.

Study of the neutral partner, $\nu_\tau$, is also of great interest.
However, the study is greatly complicated by the fact the cross section
for neutrinos interacting with matter is very small.
Nevertheless, $\nu_\tau$ has been observed for the first time.
Super-Kamiokande has also observed the evidence for $\nu_\mu$
oscillates into $\nu_\tau$.
In this paper, I will review the evidence of $\nu_\tau$ observation
and status of the $\nu_\mu \to \nu_\tau$ oscillation and then discuss the
prospects on $\nu_\tau$ oscillation experiments in the future.

There are many more results being presented at this workshop.
Space limitation precludes a comprehensive review of all the results.
Please accept my apology.

\section{Test of Lepton Universality}

In the Standard Model, the coupling of the $\tau$ lepton to the $W$ and $Z$
bosons are the same as those for the lighter leptons, $e$ and $\mu$.
The couplings to the neutral current can be studied in the decay of the $Z$ boson.
However, the couplings to the charged current can be studied in both the decay of
the $W$ boson and the decay of the $\tau$ lepton into the (virtual) $W$ boson.
This allows a test of the universality of the couplings at vastly different $Q^2$.
In this section, I will review the test of lepton university in the
charged and neutral current couplings.

\subsection{Charged Current Couplings}

Lepton universality can be tested at low $Q^2$ by comparing the measurement
of the $\tau$ lifetime ($\tau_\tau$) and leptonic
branching fractions using the relation~\cite{Marciano}:

\begin{equation}
B(\tau^- \to l^-{\bar \nu_l}\nu_\tau) = \frac{G^2_{\tau l}m^5_\tau \tau_\tau}{192\pi^3}
                                        f(\frac{m^2_l}{m^2_\tau})(1+\delta)
\end{equation}

\noindent
where the phase space factor is given by

\begin{equation}
f(x) = 1 - 8x + 8x^3 - x^4 -12x \ln x
\end{equation}

\noindent
and is 0.9726 for the muon final state;
$\delta = -0.4\%$ is the electroweak correction;
$G_{\tau l}$ is the coupling:

\begin{equation}
G_{\tau l} = \frac{g_\tau g_l}{4\sqrt{2}m^2_W} = G_F\ ,
\end{equation}

\noindent
where $m_l$, $m_\tau$, and $m_W$ are the mass of the light lepton,
$\tau$ lepton, and $W$ boson.

Both L3 and ALEPH have reported new measurements of the
leptonic branching fractions and L3 and DELPHI have reported
new measurements of the $\tau$ lifetime.
These new results can be combined~\cite{Robertson} with other
measurements to compute new averages for lifetime and leptonic
branching fractions so that they can be compared with the corresponding
measurements for an analogous decay $\mu^- \to e^-{\bar \nu_e}\nu_\mu$
to test the lepton universality.
The results on the ratios of couplings are:
\begin{eqnarray*}
\frac{g_\tau}{g_\mu} = 0.9994 \pm 0.0023\\
\frac{g_\tau}{g_e} = 1.0000 \pm 0.0023
\end{eqnarray*}

\noindent
Both ratios are consistent with the Standard Model expectation to a
precision of better than one quarter of a percent, a remarkable achievement.

\begin{figure}[htb]
\psfig{file=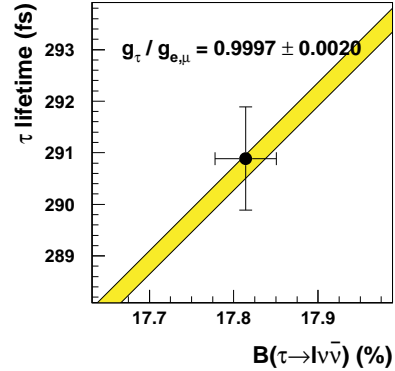,width=4.5in,angle=0}
\vspace{-2.8in}
\caption{The $\tau$ lifetime vs leptonic branching fraction.}
\label{universality}
\end{figure}

We can also compare the coupling of $\tau$ to the average coupling
for the light leptons with the assumption of $e$-$\mu$ universality.
This is a reasonable assumption since we expect any deviation from
lepton universality to occur in the heavier lepton.
The measured branching fractions for $\tau^- \to e^-{\bar \nu_e}\nu_\tau$
and $\tau^- \to \mu^-{\bar \nu_\mu}\nu_\tau$ are used to compute an
effective branching fraction for the $\tau$ decay into massless leptons,
after correcting for phase space.
Figure~\ref{universality} shows the $\tau$ lifetime vs the effective
leptonic branching fraction.
The Standard Model expectation as given by Eq.~1 is also shown as a
diagonal line with the width given by the uncertainty in the $\tau$ mass.
It is evident that the $\tau$ mass uncertainty is not negligible compared
to the errors on the lifetime and leptonic branching fraction.
The ratio of couplings is
\begin{eqnarray*}
\frac{g_\tau}{g_{e,\mu}} = 0.9997 \pm 0.0020
\end{eqnarray*}

\noindent
This is consistent with the Standard Model expectation of lepton universality.

Lepton universality can be tested at high $Q^2$ by comparing the
branching fractions of $W$ into various leptons.
The ratio of the couplings to different leptons are related to
the branching fractions by:

\begin{equation}
\frac{g_\tau}{g_{l}} = \left [\frac{B(W^- \to \tau^-\bar\nu_\tau)}
                                   {B(W \to l^-\bar\nu_l)}\right ]^{\frac{1}{2}}\ .
\end{equation}

\noindent
The measured branching fractions as averaged over the four LEP
experiments at the $e^+e^-$ collider at CERN yields~\cite{Bella}:
\begin{eqnarray*}
\frac{g_\tau}{g_\mu} = 1.022 \pm 0.016\\
\frac{g_\tau}{g_e} = 1.021 \pm 0.016
\end{eqnarray*}

For comparison, the measurements averaged over CDF and D0
experiments at the $p\bar p$ collider at Fermilab are~\cite{PDG}:
\begin{eqnarray*}
\frac{g_\tau}{g_\mu} = 0.996 \pm 0.024\\
\frac{g_\tau}{g_e} = 0.988 \pm 0.021
\end{eqnarray*}

The two sets of measurements with very different systematic errors
are consistent with each other.
Averaging over the two colliders yields:
\begin{eqnarray*}
\frac{g_\tau}{g_\mu} = 1.014 \pm 0.013\\
\frac{g_\tau}{g_e} = 1.009 \pm 0.013
\end{eqnarray*}

\noindent
The two coupling ratios are consistent with unity as expected
from lepton universality.
The precision of the measurements from the real $W$ decays
is about a factor of six less than those from the $\tau$
decay into the virtual $W$.
However, the two measurements probe lepton universality at
very different $Q^2$.
In summary, lepton universality is respected from $Q^2$ of 3 to 6500 GeV$^2$.

\subsection{Neutral Current Couplings}

Lepton universality implies that the vector and axial-vector couplings of all
leptons to the $Z$ boson are identical.
The couplings can be measured~\cite{Reid} from the leptonic partial
widths, the forward-backward asymmetry, the $\tau$ polarization, 
the $\tau$ polarization asymmetry and, in the case of polarized beam at SLC,
left-right asymmetry and left-right forward-backward asymmetry.

The leptonic partial widths of the $Z$ boson is related to the
vector ($g_{V_l}$) and axial-vector ($g_{A_l}$) couplings by:

\begin{equation}
\Gamma_{ll} = \frac{G_F m_Z^3}{6\pi\sqrt{2}} (g^2_{V_l} + g^2_{A_l})\ ,
\end{equation}

\noindent
where $m_Z$ is the mass of the $Z$ boson.
The partial width is therefore sensitive to the quadratic sum of the
coupling constants.

The parity-violating forward-backward asymmetry in the angular
distribution of the final state leptons is given by

\begin{equation}
A^{FB} = \frac{3}{4}A_l A_e\ ,
\end{equation}

\noindent
with the asymmetry parameter related to the vector and axial-vector couplings via:
\begin{equation}
A_{l} = \frac{2g_V g_A}{g^2_{V_l} + g^2_{A_l}}
      = \frac{2(\frac{g_V}{g_A})}{1 + (\frac{g_V}{g_A})^2}\ .
\end{equation}

\noindent
The asymmetry parameter is therefore sensitive to the ratio of couplings.

The final state leptons from the $Z$ decay are polarized, with a
polarization that depends on the scattering angle with respect to the beams,

\begin{equation}
P_{l}(\cos\theta) \equiv \frac{\sigma_+ - \sigma_-}{\sigma_+ + \sigma_-}
                  = \frac{\langle P_l\rangle + \frac{8}{3}A^{FB}_{pol}\frac{\cos\theta}{1+\cos^2\theta}}
                         {1   + \frac{8}{3}A^{FB}\frac{\cos\theta}{1+\cos^2\theta}}\ ,
\end{equation}

\noindent
where $\sigma_{+(-)}$ is cross section for producing a lepton of
positive (negative) helicity, $\langle P_l\rangle = -A_l $ is the
polarization averaged over all angles, and $A^{FB}_{pol} = -\frac{3}{4}A_e$
is the forward-backward polarization asymmetry.
The polarization can be measured for the case where the final state lepton
is the $\tau$ lepton by analyzing the distortions in the angular and
momentum distributions of the decay products.

At SLC, the electron asymmetry parameter $A_e$ is measured from
the left-right asymmetry in the cross sections for both left-
and right-handed electron beams:

\begin{equation}
A_{LR} = \frac{\sigma_L - \sigma_R}{\sigma_L + \sigma_R} = A_e\ .
\end{equation}

\noindent
The asymmetry parameters for the three leptons can be measured
from left-right forward-backward asymmetry:

\begin{equation}
A^{FB}_{LR} = \frac{(\sigma_{LF} - \sigma_{LB}) - (\sigma_{RF} - \sigma_{RB})}
               {\sigma_{LF} + \sigma_{LB} + \sigma_{RF} + \sigma_{RB}} = \frac{3}{4}A_l
\end{equation}

\begin{figure}[htb]
\hspace{-.8in}\psfig{file=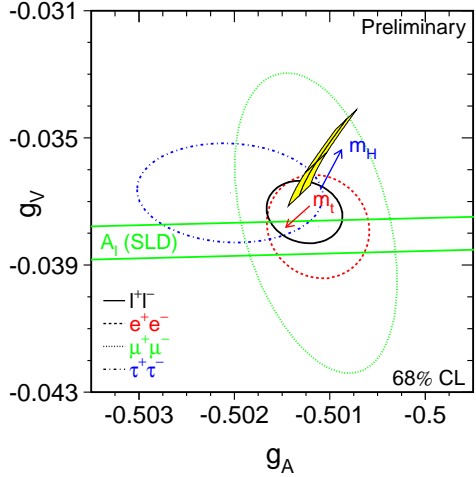,width=3.5in,angle=0}
\vspace{-.38in}
\caption{The vector vs axial-vector couplings of the leptons from a combined fit
of the LEP data.  Also shown is the result from the measurement of $A_{LR}$ from SLD.
The solid contour is from a combined fit of the LEP and SLD data.
The Standard Model expectation is shown as a
grid for $m_t = 174.3 \pm 5.1~\rm GeV$ and $m_H = 300^{+700}_{-205}~\rm GeV$.}
\label{gva}
\end{figure}

The LEP Electroweak Working Group~\cite{EWWG} has combined the results from
ALEPH, DELPHI, L3, OPAL, and SLD to extract the vector and axial-vector couplings
of the three leptons to the $Z$ boson as shown in Fig.~\ref{gva}.
The three contours overlap with each other as expected from lepton universality.
The results are also consistent with the Standard Model expectation, which
depends on the Higgs and top masses.
The ratios of the couplings are:
\begin{eqnarray*}
\frac{g^\mu_V}{g^e_V} = 0.962 \pm 0.063\ \ \ \\
\frac{g^\tau_V}{g^e_V} = 0.958 \pm 0.029\ \ \ \\
\frac{g^\mu_A}{g^e_A} = 1.0002 \pm 0.0014\\
\frac{g^\tau_A}{g^e_A} = 1.0019 \pm 0.0015
\end{eqnarray*}

The precision of the measurements from the $\tau$ lepton is better
than that for the muon because of the possibility of measuring
the polarization from the $\tau$ decay products.
The ratios are all consistent with unity as expected from lepton universality.

\section{Lorentz Structure}

The most general, local, derivative-free, lepton-number conserving,
Lorentz invariant four-lepton interaction for the leptonic decay
$\tau^- \to l^-\bar\nu_l\nu_\tau$ is given by:

\begin{equation}
{\cal M} = \frac{4G_{\tau l}}{\sqrt{2}} \sum_{\gamma = S,V,T} \sum_{i,j=L,R}
    g^\gamma_{ij}(\bar l_i \Gamma^\gamma\nu_l)(\bar\nu_\tau\Gamma_\gamma\tau_j)\ ,
\end{equation}

\noindent
where $\Gamma^\gamma$ represents the different types of charged currents:
scalar, vector, and tensor.
There are twelve complex coupling constants, $g^\gamma_{ij}$, of which two,
$g^T_{LL}$ and $g^T_{RR}$ are zero.
Since a common phase is arbitrary, there are therefore nineteen independent
real parameters which could be different for each leptonic decay.
In the Standard Model, $g^V_{LL} = 1$ for the V-A left-handed current of
the $W_L^-$ boson and all the other couplings are zero.
In many extensions to the Standard Model, additional interactions can
modify the Lorentz structure of the couplings.
For example, there can be couplings to a right-handed $W_R^-$ of the
left-right symmetric models or to the scalar currents such as those
mediated by the charged Higgs of the Minimum Supersymmetric
Standard Model (MSSM).

The coupling constants can be measured from the energy spectra of the
daughter charged leptons from the $\tau$ decay.
Integrating over the two unobserved neutrinos, the scaled energy spectrum
of the daughter lepton is:
\begin{eqnarray}
\nonumber
\lefteqn{\frac{1}{\Gamma}\frac{d\Gamma}{d x d\cos\theta}
= \frac{x^2}{2}\times\hspace{1.4in}}\\
\nonumber
& & \left \{ 12(1-x) + \frac{4\rho}{3}(8x-6)
  + 24\eta\frac{m_l}{m_\tau}\frac{1-x}{x} \right. \\
& & \left. ~\pm P_\tau\xi\cos\theta\left [ 4(1-x) + \frac{4}{3}\delta(8x-6)\right ] \right \}
\end{eqnarray}

\noindent
where $P_\tau$ is the average $\tau$ polarization, $\theta$ is the angle
between the $\tau$ spin and the daughter charged lepton momentum in the
$\tau$ rest frame, and $x = E_l/E_{max}$ is daughter charged lepton
energy scaled to the maximum energy
$E_{max} = (m_\tau^2 + m_l^2)/2m_\tau$ in the $\tau$ rest frame.

The Michel parameters~\cite{Michel} are bilinear combinations of the
coupling constants from Eq.~(11):
\begin{eqnarray}
\nonumber
\lefteqn{\rho = \frac{3}{16}(4|g^V_{LL}|^2 + 4|g^V_{RR}|^2 + |g^S_{LL}|^2 + |g^S_{RR}|^2}\\
& & + |g^S_{RL}-2g^T_{RL}|^2 + |g^S_{LR}-2g^T_{LR}|^2)
\end{eqnarray}
\begin{eqnarray}
\nonumber
\lefteqn{\eta = \frac{1}{2}Re(6g^V_{LR}g^{T*}_{LR} + 6g^V_{RL}g^{T*}_{RL} + g^S_{RR}g^{V*}_{LL}}\\
& & + g^S_{RL}g^{V*}_{LR} + g^S_{LR}g^{V*}_{RL} + g^S_{LL}g^{V*}_{RR})
\end{eqnarray}
\begin{eqnarray}
\nonumber
\lefteqn{\xi = -\frac{1}{4}(|g^S_{RR}|^2 + |g^S_{LR}|^2 - |g^S_{RL}|^2 - |g^S_{LL}|^2)}\\
\nonumber
& & +5(|g^T_{LR}|^2 - |g^T_{RL}|^2)\\
\nonumber
& & -(|g^V_{RR}|^2-3|g^V_{LR}|^2 + 3|g^V_{RL}|^2 - |g^V_{LL}|^2)\\
& & +4Re(g^S_{RL}g^{T*}_{RL} - g^S_{LR}g^{T*}_{LR})
\end{eqnarray}
\begin{eqnarray}
\nonumber
\lefteqn{\xi\delta = \frac{3}{16}(4|g^V_{LL}|^2 - 4|g^V_{RR}|^2 + |g^S_{LL}|^2 - |g^S_{RR}|^2}\\
& & + |g^S_{RL}-2g^T_{RL}|^2 - |g^S_{LR} - 2g^T_{LR}|^2)
\end{eqnarray}

\noindent
In the Standard Model, the values of the Michel parameters are:
\begin{eqnarray*}
\eta = 0\hspace{0.3in}
\rho = \frac{3}{4}\hspace{0.3in}
\xi = 1\hspace{0.3in}
\delta = \frac{3}{4} 
\end{eqnarray*}

The spectrum shape parameter $\rho$ can be measured from the momentum
spectrum of the daughter charged lepton.
The low-energy parameter $\eta$ can only be measured in the $\tau$
decay to muon because of the factor $m_l/m_\tau$ in Eq.~(12)
for the helicity flipping of the daughter charged lepton.
This parameter also can affect the leptonic decay widths and
hence can be constrained from the measured lifetime and leptonic
branching fractions using lepton universality.
On the other hand, the parameters $\xi$ and $\xi\delta$ can only
be measured for a non-zero $\tau$ polarization.
At the $Z$ resonance, the small natural polarization
($P_\tau \sim 0.14$) provides some sensitivity.
The sensitivity can be enhanced with a polarized beam as the case at SLC.
The sensitivity can also be greatly improved using the spin-spin
correlation of the $\tau$ leptons, with the opposite $\tau$ decay
as a polarization analyzer.

\begin{figure}
\psfig{file=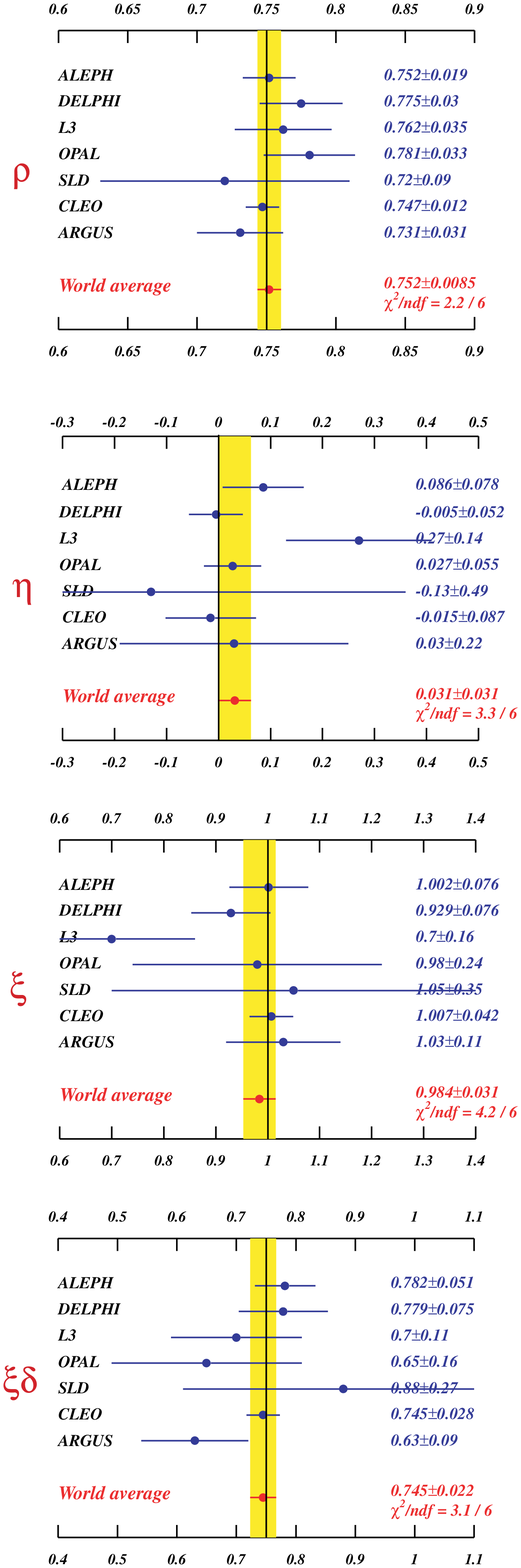,width=2.5in,angle=0}
\vspace{-.48in}
\caption{Measurements of the Michel parameters with the assumption
of $e/\mu$ universality.
The vertical lines show the Standard Model expectations.}
\label{Michel4}
\end{figure}

\begin{figure}[htb]
\psfig{file=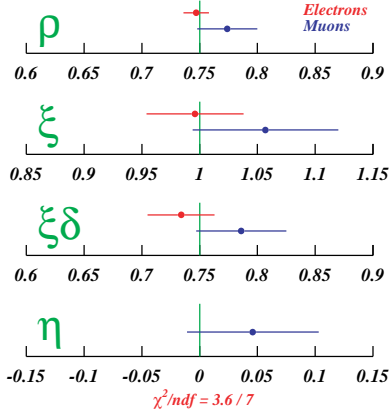,width=2.0in,angle=0}
\vspace{-.28in}
\caption{Average measurements of the Michel parameters from
the decays $\tau^- \to e^-\bar\nu_e\nu_\tau$ and
$\tau^- \to \mu^-\bar\nu_\mu\nu_\tau$.
The vertical lines show the Standard Model expectations.}
\label{Michel}
\end{figure}

The result~\cite{Boyko} on the Michel parameters extracted with the
assumption of $e/\mu$ universality is shown in Fig.~\ref{Michel4}.
All the measurements are consistent with each other as indicated
by the $\chi^2$ per degree of freedom.
The measurements are also consistent with the Standard Model expectations.
The limit on the mass of the charged Higgs extracted from the $\eta$
measurement is not competitive compared with the direct search.
Since the CLEO result on the Michel parameters is quite competitive
compared with the LEP/SLC experiments, one can expect significant
improvements in the measurements in the near future from the $b$ factories.
The Michel parameters averaged over all experiments without
the assumption of $e/\mu$ universality are shown in Fig.~\ref{Michel}.
The results are consistent with lepton universality
for a $\chi^2$ of 3.6 for 7 degrees of freedom.

\begin{figure}[htb]
\psfig{file=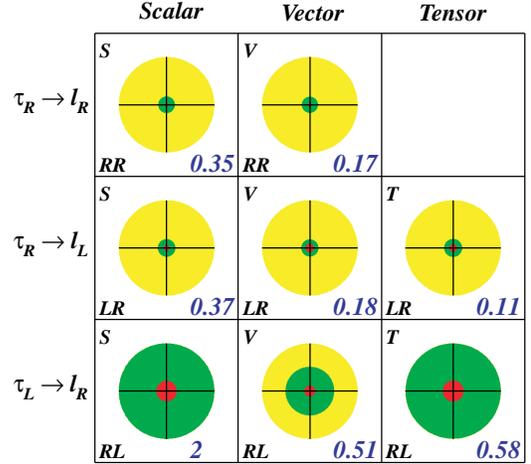,width=3.5in,angle=0}
\vspace{-.88in}
\caption{90\% CL limits (dark gray circles) in the complex plane
of the $\tau$ decay couplings for scalar, vector, and tensor
interactions of various helicities, assuming $e/\mu$ universality.
The upper limits are also indicated by the numbers on the lower
right hand corners of each box.
The light gray unit circles show the theoretically allowed areas.
The black circles show the limits for muon.}
\label{coupling}
\end{figure}

The measured Michel parameters can be used to extract limits on the
absolute values of the coupling constants $g^\gamma_{ij}$ for the
three types of interactions as shown in the complex planes in Fig.~\ref{coupling}.
Most of the measurements impose significant limits on the
coupling constants involving right-handed current $W^-_R$
as evident from the small areas shaded in dark gray over
the light gray circles, the theoretically allowed regions.
There are no limits on the ``LL'' couplings because the scalar and
vector interactions can only be distinguished in the reaction of
inverse $\tau$ decay, $\nu_\tau e \to \tau \nu_e$.
The limits from the muon are significantly more stringent.
As stated above, we can expect significant improvements in
sensitivity for new interactions from the $\tau$ decay in
the near future from $b$ factories.

\section{Hadronic Decays}

The $\tau$ lepton can decay into many hadronic states due to its large mass.
The hadronic decay products have distinctive charge conjugation $(C)$ and
isospin (and hence $G$-parity) signatures, a reflection of the quantum number
of the charged hadronic weak current.
The weak current is classified according to its $G$-parity:

\vspace{0.1in}
\noindent $
\begin{array}{lll}
{\rm Vector:} & G = +1, J^P = 1^-      & e.g.\ \rho^-\\
{\rm Axial:}  & G = -1, J^P = 0^-, 1^+ & e.g.\ \pi^-, a^-_1
\end{array}$
\vspace{0.1in}

\noindent
These are known as the first class currents.
Currents with opposite $G$-parity are called the second class
currents~\cite{Weinberg} and are suppressed by the order $\alpha^2$
or $10^{-4}$ in the Standard Model.
Examples of second class current decays are $\tau^- \to a^-_0\nu_\tau$
and $\tau^- \to b^-_1\nu_\tau$.

The large number of hadronic channels allow the study of many aspects
of the Standard Model, including
test of the Conserved-Vector-Current (CVC) hypothesis~\cite{Feynman},
isospin symmetry, QCD sum rules, study of light meson resonances, and
measurements of the strong coupling constant, strange quark mass, and
hadronic contribution of the muon anomalous magnetic moment.
In this section, I will review the topics in which there are significant
progress since Tau98.

\subsection{Test of CVC}

The CVC hypothesis relates the coupling strength of the weak charged vector
current in the $\tau$ decay to the electromagnetic neutral vector current
in $e^+e^-$ annihilation.
This allows the calculation of the branching fraction for the $\tau$ decays
into the vector final state from the $e^+e^-$ cross section for the
corresponding isovector final state ($I = 1$).
For example, the branching fraction for $\tau^- \to \pi^-\pi^0\nu_\tau$ is
given by:
\begin{eqnarray}
\nonumber
\lefteqn{B_{\pi\pi} = \frac{G^2_F\tau_\tau\cos^2\theta_c}{128\pi^4\alpha^2 m_\tau^3}
                      \int^{m_\tau^2}_{4m_\pi^2} dq^2q^2}\\
& & (m_\tau^2-q^2)^2(m_\tau^2+2q^2) \sigma_{e^+e^- \to \pi^+\pi^-}^{I=1}(q^2)
\end{eqnarray}

\noindent
where $q$ is the center-of-mass energy of the $e^+e^-$ collision,
$m_\pi$ is the $\pi$ mass, $\theta_c$ is the Cabbibo angle, and
$\alpha$ is the fine structure constant.

At Tau98, the CVC prediction~\cite{Eidelman98} for $B_{\pi\pi}$ was:
\begin{eqnarray*}
B_{\pi\pi}^{CVC} = (24.52 \pm 0.33)\%
\end{eqnarray*}
This is to be compared with the current world average~\cite{PDG98} of
\begin{eqnarray*}
B_{\pi\pi} = (25.31 \pm 0.18)\%\ .
\end{eqnarray*}
\noindent
The fractional difference between the two branching fractions is
$(3.1 \pm 1.5)\%$, which is slightly over $2\sigma$ (standard deviation).

At this Workshop, there is a new measurement~\cite{Chen} by L3
on the branching fraction for $\tau^- \to h^-\pi^0\nu_\tau$ of
$(25.38 \pm 0.18 \pm 0.14)\%$.
\noindent
Correcting for the kaon contribution~\cite{PDG} of
$(0.449 \pm 0.034)\%$ yields:
\begin{eqnarray*}
B_{\pi\pi} = (24.93 \pm 0.23)\%\ .
\end{eqnarray*}
This is consistent with the world average.
Including this branching fraction into the world average yields a
new world average of
\begin{eqnarray*}
B_{\pi\pi} = (25.17 \pm 0.14)\%\ .
\end{eqnarray*}
This is to be compared with the new CVC expectation~\cite{Eidelman}
presented at this Workshop based on new data on $e^+e^- \to \pi^+\pi^-$:
\begin{eqnarray*}
B_{\pi\pi}^{CVC} = (24.94 \pm 0.23)\%\ .
\end{eqnarray*}
The fractional difference between the two branching fractions is
$(0.9 \pm 1.1)\%$ and hence there is good agreement between the
measured branching fraction and the CVC prediction.

\begin{figure}[htb]
\psfig{file=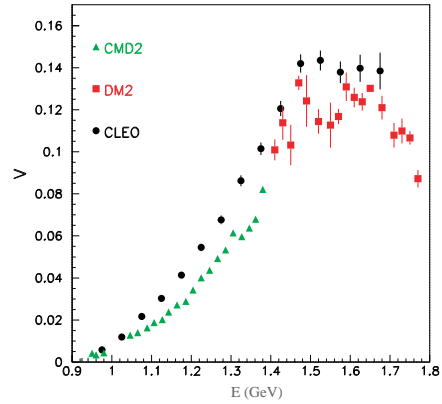,width=2.5in,angle=0}
\vspace{-.38in}
\caption{Spectral function of the decay
$\tau^- \to 2\pi^-\pi^+\pi^0\nu_\tau$.}
\label{4pi}
\end{figure}

The agreement between the measured branching fraction for the $4\pi$
final state and the CVC expectation is not as good.
Figure~\ref{4pi} shows the spectral functions~\cite{Eidelman}
for the $4\pi$ final state extracted from
$\tau^- \to 2\pi^-\pi^+\pi^0\nu_\tau$ by CLEO and from
$e^+e^- \to 2\pi^+2\pi^-$ and $\pi^+\pi^-2\pi^0$ by CMD2 and DM2.
The recent CMD2 data from the VEPP-2M collider has significantly
better precision than the vintage DM2 data.
The spectral function extracted from the $\tau$ decay is consistently
higher across the entire energy range.
Since branching fraction is related to the integral of the spectra
function, the discrepancy is also expected between the measured
branching fraction~\cite{4pi} and the CVC prediction~\cite{Eidelman}:
\begin{eqnarray*}
& B_{3\pi\pi^0} = (4.19 \pm 0.10 \pm 0.21)\%\\
& B_{3\pi\pi^0}^{CVC} = (3.55 \pm 0.20)\%
\end{eqnarray*}
The discrepancy is $\sim 18\%$ with a statistical significant
of $\sim 2\sigma$.

\begin{figure}[htb]
\psfig{file=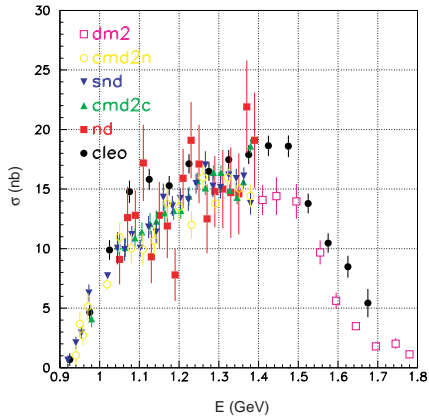,width=2.5in,angle=0}
\vspace{-.38in}
\caption{Cross section for the process $e^+e^- \to \pi^0\omega$.}
\label{piomega}
\end{figure}

There is also a discrepancy in a subset of the $4\pi$ final states:
$\tau^- \to \pi^-\omega\nu_\tau$ and $e^+e^- \to \pi^0\omega$.
Instead of comparing the extracted spectral functions, the $\tau$ branching
fraction is converted into the cross section for $e^+e^- \to \pi^0\omega$
as shown in Fig.~\ref{piomega}.
There is good agreement between various experiments at low energy
but at high energy, the cross section extracted from the $\tau$ decay
is significantly higher than that of DM2, indicating a normalization
problem in the DM2 data.
The discrepancy is also reflected in the measured branching
fraction~\cite{PDG} and the CVC prediction~\cite{Eidelman}:
\begin{eqnarray*}
& B_{\pi\omega} = (1.92 \pm 0.07)\%\\
& B_{\pi\omega}^{CVC} = (1.73 \pm 0.06)\%
\end{eqnarray*}
As expected, the magnitude of the discrepancy is somewhat smaller,
$\sim 10\%$, but the statistical significant remains the same,
$\sim 2\sigma$.
It should be noted that $B_{\pi\omega}$ includes a small contribution
from $\tau^- \to K^-\omega\nu_\tau$ which is both phase-space and
Cabbibo-suppressed.

\subsection{Measurement of $\rho(1450)$ Parameters}

The hadronic $\tau$ decay provides a clean environment to study the
light meson resonance parameters.
The CLEO experiment~\cite{4pi} has extracted the $\rho(1450)$
parameters from a fit to the $\pi\omega$ mass spectrum in the
decay $\tau^- \to \pi^-\omega\nu_\tau$:
\begin{eqnarray*}
M_{\rho'} = 1523 \pm 10 \ \rm MeV\\
\Gamma_{\rho'} = 400 \pm 35 \ \rm MeV
\end{eqnarray*}
The mass is significantly different from the fit to the mass
spectrum~\cite{2pi} in the decay $\tau^- \to \pi^-\pi^0\nu_\tau$
although the width is consistent:
\begin{eqnarray*}
M_{\rho'} = 1406 \pm 15 \ \rm MeV\\
\Gamma_{\rho'} = 455 \pm 41 \ \rm MeV
\end{eqnarray*}
The parameters are also different from those extracted by PDG~\cite{PDG}
from a combination of data from fixed target experiments,
$e^+e^- \to \pi^+\pi^-$ and $e^+e^- \to \eta\pi^+\pi^-$, as well as
from earlier results from the decay $\tau^- \to \pi^-\pi^0\nu_\tau$:
\begin{eqnarray*}
M_{\rho'} = 1465 \pm 25 \ \rm MeV\\
\Gamma_{\rho'} = 310 \pm 60 \ \rm MeV
\end{eqnarray*}
The origin of these differences is unknown.

\subsection{Test of CVC and Isospin Symmetry in Six-Pion Decays}

The CLEO experiment~\cite{6pi} has presented new results
on two six-pion decays, $\tau^-\to 2\pi^-\pi^+3\pi^0\nu_{\tau}$
and $\tau^-\to 3\pi^-2\pi^+\pi^0\nu_{\tau}$.
There is no experimental information on the decay
$\tau^-\to \pi^-5\pi^0\nu_{\tau}$ due to the difficulty in extracting
a signal from the large combinatoric background.
The decays may proceed through the $\rho$, $\omega$ or $\eta$ intermediate states.
The six-pion decays are therefore a mixture of vector
and axial-vector current decays.
The contribution from the axial-vector current decay,
$\tau^-\to (3\pi)^-\eta\nu_{\tau}$, must be subtracted in the test
of CVC and isospin symmetry.

The new results on the six-pion decay branching fractions are:
\begin{eqnarray*}
B(\tau^-\to 2\pi^-\pi^+3\pi^0\nu_{\tau})=(2.2 \pm 0.3 \pm 0.4)\times 10^{-4}\\
B(\tau^-\to 3\pi^-2\pi^+\pi^0\nu_{\tau})=(1.7 \pm 0.2 \pm 0.2)\times 10^{-4}
\end{eqnarray*}
The results represent significant improvement in precision over
previous measurements~\cite{PDG}.

The experiment has also searched for $\eta$ or $\omega$ intermediate states.
For the decay $\tau^-\to 2\pi^-\pi^+3\pi^0\nu_{\tau}$, the branching
fractions with the intermediate decays $\eta \to 3\pi^0$,
$\eta \to \pi^+\pi^-\pi^0$ and $\omega \to \pi^+\pi^-\pi^0$ are
\begin{eqnarray*}
B(\tau^-\to 2\pi^-\pi^+\eta\nu_{\tau})=(2.9 \pm 0.7 \pm 0.5)\times 10^{-4}\\
B(\tau^-\to \pi^-2\pi^0\eta\nu_{\tau})=(1.5 \pm 0.6 \pm 0.3)\times 10^{-4}\\
B(\tau^-\to \pi^-2\pi^0\omega\nu_{\tau})=(1.5 \pm 0.4 \pm 0.3)\times 10^{-4}
\end{eqnarray*}
In the decay $\tau^-\to 3\pi^-2\pi^+\pi^0\nu_{\tau}$, the branching
fractions for the intermediate decays $\eta \to \pi^+\pi^-\pi^0$
and $\omega \to \pi^+\pi^-\pi^0$ are
\begin{eqnarray*}
B(\tau^-\to 2\pi^-\pi^+\eta\nu_{\tau})=(1.9 \pm 0.4 \pm 0.3)\times 10^{-4}\\
B(\tau^-\to 2\pi^-\pi^+\omega\nu_{\tau})=(1.2 \pm 0.2 \pm 0.1)\times 10^{-4}
\end{eqnarray*}
This constitutes the first observation of the decay
$\tau^-\to 2\pi^-\pi^+\omega\nu_{\tau}$.
The results are consistent with saturation of the two six-pion decays
by $\eta$ and $\omega$ intermediate states.
The branching fractions for the two decays with
$\omega$ in the final states are somewhat smaller than the  
recent calculations by Gao and Li~\cite{GaoAndLi}:
\begin{eqnarray*}
B(\tau^-\to \pi^-2\pi^0\omega\nu_{\tau})=2.16\times 10^{-4}\\
B(\tau^-\to 2\pi^-\pi^+\omega\nu_{\tau})=2.18\times 10^{-4}
\end{eqnarray*}

The results on the six-pion decays can be compared with the isospin
symmetry and CVC predictions, after correcting for the contributions
from the axial-vector current $\tau^- \to (3\pi)^-\eta \nu_{\tau}$,
which also violates isospin conservation with the decays
$\eta \to 3\pi^0$ and $\pi^+\pi^-\pi^0$.
To reduce the uncertainty in the corrections, CLEO uses measurements
from these decays and the decay $\eta \to \gamma\gamma$~\cite{3pieta}
to obtain the average branching fractions:
\begin{eqnarray*}
{\bar B}(\tau^-\to 2\pi^-\pi^+\eta\nu_{\tau})=(2.4\pm0.5)\times 10^{-4}\\
{\bar B}(\tau^-\to \pi^-2\pi^0\eta\nu_{\tau})=(1.5\pm0.5)\times 10^{-4}
\end{eqnarray*}
This yields the vector current branching fractions:
\begin{eqnarray*}
B_V(\tau^-\to 2\pi^-\pi^+3\pi^0\nu_{\tau})=(1.1 \pm 0.4)\times 10^{-4}\\
B_V(\tau^-\to 3\pi^-2\pi^+\pi^0\nu_{\tau})=(1.1 \pm 0.2)\times 10^{-4}
\end{eqnarray*}
corresponding to $\sim$50\% and $\sim$65\%, respectively, of the inclusive
six-pion branching fractions.

\begin{figure}[htb]
\centering
\centerline{\hbox{\psfig{figure=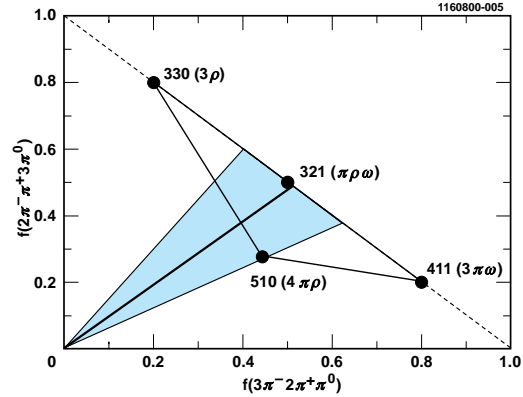, width=7.0cm}}}
\vspace{-0.2in}
\caption{Decay fractions of $\tau^-\to (6\pi)^-\nu_{\tau}$ as measured by CLEO.
         The thick solid line through the origin represents the measurement.
         The shaded area indicates the one standard deviation region,
         calculated with correlated errors taken into account.
         The triangle bounded by the dots shows the isospin expectation.}      
\label{figure:isospin}
\end{figure}

The isospin model~\cite{Pais} classifies $n$-pion final states into
orthogonal isospin partitions and determines their contributions 
to the branching fractions.
The partitions are labeled by three integers $(n_1,n_2,n_3)$,
where $n_3$ is the number of isoscalar subsystems of three pions,
$n_2-n_3$ is the number of isovector systems of two pions, and $n_1-n_2$
is the number of single pions. For  $n=6$ there are four partitions:
510 ($4\pi\rho$), 330 ($3\rho$), 411 ($3\pi\omega$), and 321
($\pi\rho\omega$), denoted according to the lowest mass states.
The isospin model imposes constraints on the relative branching
fractions, which can be tested by comparing the following two fractions
\begin{eqnarray}
{\rm f}_{2\pi^-\pi^+3\pi^0} = \frac{B_V(\tau^-\to 2\pi^-\pi^+3\pi^0\nu_\tau)}
                     {B_V(\tau^-\to (6\pi)^-\nu_\tau)}\\
{\rm f}_{3\pi^-2\pi^+\pi^0} = \frac{B_V(\tau^-\to 3\pi^-2\pi^+\pi^0\nu_\tau)}
                     {B_V(\tau^-\to (6\pi)^-\nu_\tau)}
\end{eqnarray}
where $B_V(\tau^-\to (6\pi)^-\nu_\tau)$
is the sum of the branching fractions for the three six-pion vector decays.
Figure~\ref{figure:isospin} shows ${\rm f}_{2\pi^-\pi^+3\pi^0}$
vs.~${\rm f}_{3\pi^-2\pi^+\pi^0}$ with the new measurement of
the branching fractions.
The measurement is presented as a line because
$B_V(\tau^-\to \pi^-5\pi^0\nu_\tau)$ has not been measured yet.
The result is consistent with the isospin expectation since
the experimental measurement overlaps with the isospin triangle.
The result indicates the 321
($\pi\rho\omega$) partition is dominant because the decays
$\tau^-\to \pi^-2\pi^0\omega\nu_{\tau}$ and     
$\tau^-\to 2\pi^-\pi^+\omega\nu_{\tau}$ saturate the
six-pion (vector) decays.

A significant fraction of the experimentally allowed area is
still outside the isospin allowed region.
This can only be improved if the branching fraction for the
experimentally challenging decay $\tau^-\to \pi^-5\pi^0\nu_{\tau}$
has been measured.
A similar analysis on the five-pion decays should be performed.

The measured branching fractions can be compared with the CVC
expectations~\cite{Eidelman97} based on the measured cross sections for
$e^+e^- \to 6\pi$:
\begin{eqnarray*}
B(\tau^-\to 2\pi^-\pi^+3\pi^0\nu_{\tau}) \ge (2.5 \pm 0.4)\times10^{-4}\\
B(\tau^-\to 3\pi^-2\pi^+\pi^0\nu_{\tau}) \ge (2.5 \pm 0.4)\times10^{-4}
\end{eqnarray*}
The predictions are significant larger than the measured branching fractions
for the six-pion vector decays.
The discrepancy is even more significant if we compare the
predicted inclusive branching fraction
\begin{eqnarray*}
B(\tau^-\to (6\pi)^-\nu_{\tau}) \ge (12.3 \pm 1.9)\times10^{-4}
\end{eqnarray*}
with the sum of the measured six-pion vector
branching fractions under the assumption that
$B(\tau^-\to \pi^-5\pi^0\nu_{\tau})$ is comparable with
or smaller than $B(\tau^-\to 2\pi^-\pi^+3\pi^0\nu_{\tau})$
and $B(\tau^-\to 3\pi^-2\pi^+\pi^0\nu_{\tau})$ as expected
by isospin symmetry:
\begin{eqnarray*}
B(\tau^-\to \pi^-5\pi^0\nu_{\tau}) \le
       \frac{9}{26}[ B(\tau^-\to 2\pi^-\pi^+3\pi^0\nu_{\tau})\\
+ B(\tau^-\to 3\pi^-2\pi^+\pi^0\nu_{\tau})]
\end{eqnarray*}
This assumption is consistent with the observation that the
six-pion vector decays are saturated by intermediate
states with an $\omega$ meson, which implies a small decay
width for the 510 ($4\pi\rho$) state, the only state that
contributes to the decay $\tau^-\to \pi^-5\pi^0\nu_{\tau}$.
The discrepancy might be explained by sizable presence
of $I=0$ states in the $e^+e^-$ annihilation data that should be 
subtracted before calculating the CVC prediction.

\section{Anomalous Magnetic Moment}

The g-2 experiment at CERN is currently being repeated at Brookhaven
National Laboratory with a precision that is sensitive to new physics
comparable to the electroweak corrections.
The largest uncertainty in calculating the muon anomalous magnetic
moment is the contribution from hadronic vacuum polarization.
The contribution is calculated using the measured $e^+e^-$ cross
section of hadronic final states and is dominated by
$\sigma(e^+e^- \to \pi^+\pi^-)$.
There is a new estimate of the $\pi\pi$ contribution with the
addition of the recent data from CMD-2~\cite{Eidelman}:
\begin{eqnarray*}
a_{\mu}^{\pi\pi} = (498.8 \pm 5.0)\times 10 ^{-10}\ .
\end{eqnarray*}

The hadronic decays of the $\tau$ lepton provide an alternative
estimate of the hadronic vacuum polarization with different systematic error.
There are estimates of the $\pi\pi$ contribution using the
ALEPH~\cite{Davier} and CLEO~\cite{2pi} measurements of the decay rate
for $\tau^-\to \pi^-\pi^0\nu_{\tau}$:
\begin{eqnarray*}
a_{\mu}^{\pi\pi} = (502.2 \pm 6.9)\times 10 ^{-10} & & (\rm ALEPH)\\
a_{\mu}^{\pi\pi} = (513.1 \pm 5.8)\times 10 ^{-10} & & (\rm CLEO)
\end{eqnarray*}
The two measurements are not inconsistent with each other or the
measurement using $\sigma(e^+e^- \to \pi^+\pi^-)$ given the errors.
However, the difference between the two measurements is three times
the expected precision of $4 \times 10 ^{-10}$ from the new g-2
experiment~\cite{g-2}.
If we combined the three measurements despite the possible discrepancy,
then:
\begin{eqnarray*}
a_{\mu}^{\pi\pi} = (504.3 \pm 3.3)\times 10 ^{-10}\ .
\end{eqnarray*}
The uncertainty is still not negligible in comparison with the
expected precision of the g-2 experiment.
New measurements from b-factory and other LEP experiments will
provide a powerful consistency check.
However, it should be noted that the extraction of $a_{\mu}^{\pi\pi}$
from the $\tau$ decay requires the assumption of CVC which should
be closely scrutinized because the combined $\tau$
measurement has now reached a precision of $\sim 1\%$.

\section{Strange Quark Mass}

The strange quark mass is one of the fundamental parameters of
the Standard Model.
For example, it is needed in the extraction of the CKM matrix
elements from the measurement of the direct CP violation
parameter, $\epsilon'/\epsilon$, in the kaon system.
In the $\tau$ decay, the strange quark decay rate depends on the
strange quark mass, unlike the non-strange decay, which is
insensitive to the up and down quark masses.
The strange quark mass suppresses the decay rate by $\sim 10\%$.
Extraction of the mass requires the calculation of the somewhat
controversial non-perturbative corrections.
Nevertheless, there has been significant progress in the past few
years in the technique for extracting the mass.

In principle, the strange quark mass can be extracted from the
inclusive strange decay rate.
However, a better technique is to extract the mass from the difference
between the non-strange and strange spectral moments so that the
mass-independent non-perturbative corrections cancel to first order.
Using the latest results on the branching fractions for the strange
decays by CLEO and OPAL together with the published strange spectral
functions measured by ALEPH, Davier \etal~\cite{ms} extracts a new
estimate of the mass in the $\rm \overline{MS}$ scheme:
\begin{eqnarray*}
m_s(m_\tau) = 112 \pm 23\ \rm MeV\ ,
\end{eqnarray*}
which, using four-loop running, yields a mass at a slightly higher
scale of 
\begin{eqnarray*}
m_s(\rm 2\ GeV) = 107 \pm 22\ \rm MeV\ .
\end{eqnarray*}
This can be compared with the mass extracted by Maltman and
Kambor~\cite{Maltman}
\begin{eqnarray*}
m_s(\rm 2\ GeV) = 115.1 \pm 13.6 \pm 11.8 \pm 9.7\ \rm MeV
\end{eqnarray*}
using the ALEPH data with an optimized finite energy sum rule,
where the first error is due to the experimental uncertainty, the
second due to uncertainty in the CKM matrix element $V_{us}$,
and the third due to the theoretical uncertainty.
It is reassuring to see that both techniques give consistent results.

The estimates can be compared with the result from the lattice
calculation~\cite{Lubicz}
\begin{eqnarray*}
m_s(\rm 2\ GeV) = 110 \pm 25\ \rm MeV\ .
\end{eqnarray*}
The $\tau$ result therefore provides an independent estimate of
this fundamental parameter with a very different systematic error.
With the advent of the b-factory experiments with excellent kaon
identification, we expect more precise measurements of the strange
spectral functions and hence a better determination of the strange
quark mass.

\section{Direct Measurement of Electric Dipole Moment}

There are three kinds of dipole moments from the couplings of
the $\tau$ lepton to the $\gamma$, $Z$, and $W$ bosons:
anomalous (electromagnetic) magnetic ($a_\tau$)
and electric dipole ($d_\tau$) moments, anomalous weak magnetic
($a_\tau^W$) and electric dipole ($d_\tau^W$) moments, anomalous
charged weak magnetic ($\kappa$) and electric dipole
($\tilde{\kappa}$) moments.
In the Standard Model, radiative corrections produce a non-zero
anomalous magnetic dipole moment~\cite{Samuel}
\begin{eqnarray*}
a_\tau  = \frac{g-2}{2} = (1.1773 \pm 0.0003) \times 10^{-3}\ .
\end{eqnarray*}
The electric dipole moment $d_\tau$ is zero for pointlike fermions;
a non-zero value would violate both $T$ and $P$ and hence $CP$ invariance.
The weak magnetic dipole moment is very small~\cite{Bernabeu}:
\begin{eqnarray*}
d_\tau^W = -(2.10 + 0.61i) \times 10^{-6}\ .
\end{eqnarray*}
The weak electric dipole moment is also very small but non-zero
due to CP violation in the CKM matrix:
\begin{eqnarray*}
d_\tau^W \approx 3 \times 10^{-37}\ \rm e\cdot cm\ .
\end{eqnarray*}
The charged weak magnetic and electric dipole moments are expected
to be very small.
Since all the dipole moments are expected to be small or zero,
measurement of the dipole moments provides a sensitive probe of
physics beyond the Standard Model.

The ARGUS experiment~\cite{ARGUS} has searched for $CP$ violation due
to a non-zero electric dipole moment in the production of $\tau$ pairs.
The $CP$ violation produces a charge dependent momentum correlation.
The experiment constructed optimized observables which took into
account all available information on the observed decay products.
No evidence for $CP$ violation was found, resulting in the
following 95\% CL upper limit on the electric dipole moment:
\begin{eqnarray*}
Re(d_\tau) < 4.6 \times 10^{-16}\ \rm e\cdot cm\\
Im(d_\tau) < 1.8 \times 10^{-16}\ \rm e\cdot cm
\end{eqnarray*}
These are the first direct measurement of the electric dipole moment.
For comparison, the limit extracted~\cite{L3} by the L3 experiment
from an analysis of the reaction $e^+e^- \to \tau^+\tau^-\gamma$ is
\begin{eqnarray*}
|d_\tau| < 3.1 \times 10^{-16}\ \rm e\cdot cm\ ,
\end{eqnarray*}
with the assumption of zero anomalous magnetic dipole moment.

\section{Direct Observation of the $\tau$ Neutrino}

The existence of the $\tau$ neutrino was postulated after the
discovery of the the $\tau$ lepton in 1975.
Much direct evidence for its existence has accumulated in the
intervening 25 years.
It has been assumed that $\nu_\tau$ is a sequential neutrino
to $\nu_\mu$ and $\nu_e$ in the Standard Model.
This elusive particle has finally been observed directly~\cite{DONUT}
by the DONUT experiment at Fermilab.

Evidence for $\nu_\tau$ relies on identification of the $\tau$
leptons produced in the charged current $\nu_\tau$ interactions.
The $\tau$ neutrinos are produced by directing 800 GeV/c protons
onto a tungsten dump via the decay $D_s^+ \to \tau^+\nu_\tau$
with an average energy $\langle E_{\nu_\tau}\rangle = 54~\rm GeV$.
The $e$ and $\mu$ neutrinos are also produced in the
$\mu$, $\pi$, $K$, and $D$ decays.
These neutrinos could interact via a charge or neutral current
to produce background events.

\begin{figure}[htb]
\centering
\centerline{\hbox{\psfig{figure=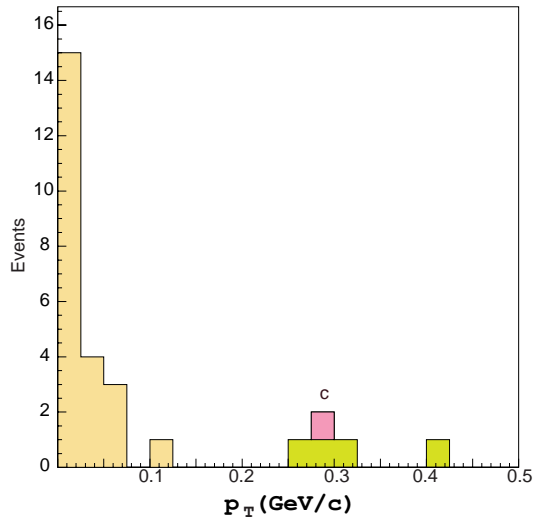, width=7.0cm}}}
\vspace{-0.2in}
\caption{Transverse momentum of kink events.
         The event labeled as c is identified as a charm background
         (see text).}      
\label{figure:pt}
\end{figure}

\begin{figure}[htb]
\centering
\centerline{\hbox{\psfig{figure=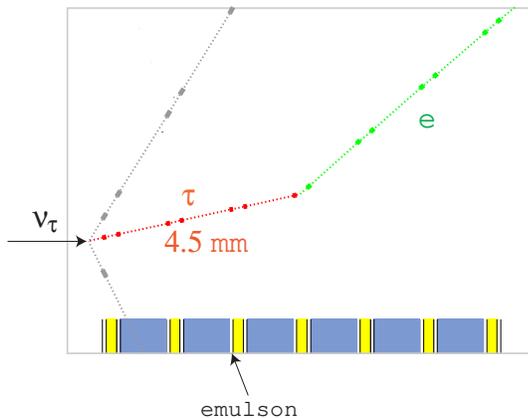, width=7.0cm}}}
\vspace{-0.2in}
\caption{A $\tau$ candidate produced by the charged current interaction
         of a $\tau$ neutrino.  The kink indicates the decay of the
         $\tau$ candidate into an electron and two neutrinos.
         The horizontal bars at the bottom show the target configuration
         in the vicinity of the primary vertex.  Dark solid bars represent
         the stainless steel plates and light grey bars represent the
         plastic substrate with emulsion (white bars) on both sides.}      
\label{figure:DONUT}
\end{figure}

Critical to the $\nu_\tau$ identification is the emulsion target
which contains $\sim 200$ plastic substrate sheets with
$\sim 100\ \mu\rm m$ of emulsion coated on both sides.
Each track leaves a trail of emulsion grains, providing the spatial
location with excellent resolution, 0.3-0.4~$\mu$m.
A track segment is constructed from the emulsion grains in each
emulsion sheet.
Combining the track segments from the emulsion sheets allows the
observation of the kink in a $\tau$ decay.
A $\nu_\tau$ candidate from a charge current interaction
could produce several tracks emanating from the (primary)
interaction vertex, including a $\tau$ candidate track with a kink.
The experiment requires both the $\tau$ candidate track and its
daughter track to have impact parameter $< 5\ \mu\rm m$, taking
advantage of the high spatial resolution of the emulsion for
background suppression.
The background includes random association of background tracks,
re-scattering of a track emanating from the primary interaction
vertex, and charm decay from a charged current interaction.
The probability of re-scattering decreases rapidly with traverse
momentum $p_t$, in contrast to $\tau$ decays in which $p_t$ peaks
at 400 MeV/c.
Figure~\ref{figure:pt} shows the $p_t$ spectrum of the candidate tracks.
There are five $\tau$ candidate tracks with $p_t >$ 250 MeV/c.
One of the high $p_t$  event is identified as a charm background
from a charged current interaction by the presence of an electron
emanating from the primary vertex.
The remaining four events are $\tau$ candidates with no evidence
for a lepton from the primary vertex.
Figure~\ref{figure:DONUT} shows a picture of one of the candidates.

The sample of four events is consistent with the expected signal
of $4.1 \pm 1.4$ events.
The total estimated background is $0.41 \pm 0.15$ events from
mis-tagged charm decays and secondary interactions.
The Poisson probability that the four events are due to the background
processes is $8 \times 10^{-4}$.
The $\tau$ neutrino has therefore been directly observed after the
first sighting of the $\tau$ lepton 25 years ago!
This completes the picture of the Standard Model of three generations
of fermions.

The DONUT experiment has proven the principle of directly detecting
the $\tau$ neutrino.
This is an important milestone for the experiments currently being
constructed to search for $\nu_\tau$ appearance to verify the signal
of $\nu_\mu \to \nu_\tau$ oscillation observed by Super-Kamiokande.
This will be reviewed in the next section.

\section{Neutrino Oscillations}

In the Standard Model, neutrinos are assumed to be massless.
However, it can also accommodate neutrinos with mass.
If neutrinos have mass, then they can mix with one another,
thereby violating lepton family number conservation.
In a two-flavor oscillation between $\nu_\mu$ and $\nu_\tau$,
for example, the probability for oscillation is given by
\begin{equation}
P(\nu_\mu \to \nu_\tau) = \sin^2(2\theta_{\mu\tau})\sin^2(\pi L/L_0)\ ,
\end{equation}
where $\theta_{\mu\tau}$ is the mixing angle, $L$ is distance traversed
by $\nu_\mu$ in meter, and the oscillation length is given by
\begin{equation}
L_0 = \frac{2.48E_\nu}{\Delta m^2}\ ,
\end{equation}
with $\Delta m^2 = m_\tau^2 - m_\mu^2$ being the difference of the
mass squares in eV$^2$ and $E_\nu$ being the energy of the neutrino in GeV.

Evidence for neutrino oscillations has been seen in the solar $\nu_e$
deficit ($\nu_e$ disappearance), atmospheric $\nu_\mu$ deficit ($\nu_\mu$
disappearance), and $\bar \nu_e$ excess in $\pi^+$ decays
($\bar\nu_\mu \to \bar\nu_e$ appearance).
It is difficult to accommodate all the apparent oscillations in a
three neutrino generation model.
However, an additional sterile neutrino ($\nu_s$) can easily explain all
the oscillations.

In this Section, I will review the new result presented at this
Workshop from Super-Kamiokande~\cite{SuperK} on the atmospheric
$\nu_\mu$ deficit and from CHORUS and NOMAD on the
search~\cite{Shibuya} for $\nu_\mu \to \nu_\tau$ oscillation.
I will then discuss the current status and future prospect
of long baseline experiments on the $\nu_\tau$ oscillation.

\subsection{Observation of Atmospheric $\nu_\mu$ Deficit}

Atmospheric neutrinos are produced by collisions of cosmic rays with
the upper atmosphere.
If there is no oscillation, the neutrino flux should be relatively
uniform with no up-down asymmetry.
The neutrinos originate from the decay chain: $\pi \to \mu\bar\nu_\mu$
with $\mu \to e\bar\nu_e\nu_\mu$.
The ratio of the muon to electron neutrino flux should be $\sim 2$ in
the no oscillation scenario.
Super-Kamiokande observed a very significant $\nu_\mu$ deficit.
This can be attributed to the oscillation of
$\nu_\mu$ to $\nu_e$, $\nu_\tau$ or $\nu_s$.
The oscillation cannot be pure $\nu_\mu \to \nu_e$ because there is
no significant excess of $\nu_e$ originated from below.
In addition, the CHOOZ~\cite{CHOOZ} and Palo Verde~\cite{Palo_Verde}
experiments have ruled out disappearance of reactor $\bar\nu_e$
for similar parameters.
The $\nu_\mu \to \nu_s$ oscillation scenario has been investigated
using two techniques based on the fact that $\nu_s$ has no neutral
current coupling.
First, there is an MSW-like effect for $\nu_s$ in the Earth that
effectively suppresses oscillation.
The effect is more prominent at higher energies and no distortion
in the angular distribution of high energy events is observed.
Second, there is no deficit of up-going neutral current events
in a neutral current enriched sample of multi-ring events.
These observations exclude the $\nu_\mu \to \nu_s$ oscillation
at the 99\% CL.

\begin{figure}
\centering
\centerline{\hbox{\psfig{figure=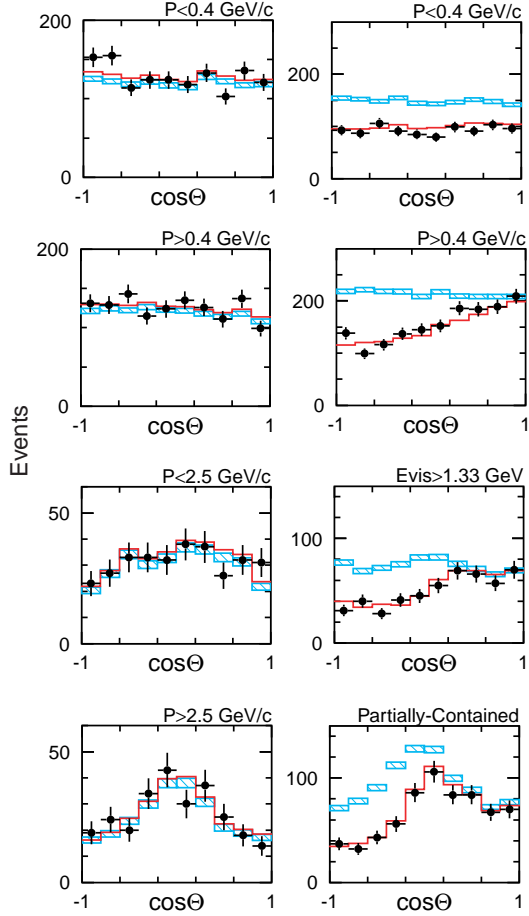, width=7.0cm}}}
\vspace{-0.2in}
\caption{Zenith angle distributions of $e$-like (left) and $\mu$-like
         (right) events of various visible energies.  The lines show
         the best fit with $\nu_\mu \to \nu_\tau$ oscillation.  The
         hatched lines indicate the expectations without oscillation.}      
\label{figure:zenith}
\end{figure}

\begin{figure}[htb]
\centering
\centerline{\hbox{\psfig{figure=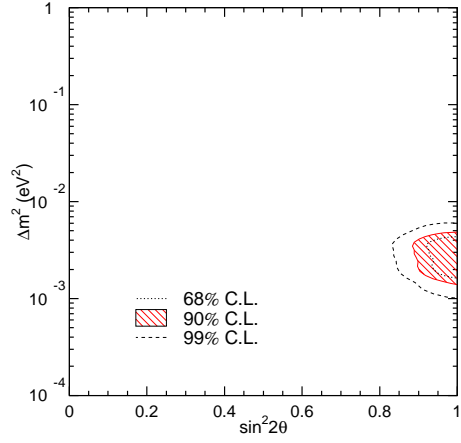, width=6.0cm}}}
\vspace{-0.2in}
\caption{The 68, 90, and 99\% confidence intervals on the
         $\nu_\mu \to \nu_\tau$ oscillation parameters.}      
\label{figure:oscillation}
\end{figure}

The zenith angle dependent of the atmospheric neutrino events of
various visible energies is shown in Fig.~\ref{figure:zenith}.
For the $e$-like events, events most likely produced by $\nu_e$
interactions, there is no evidence of any deficit or excess.
However, for the $\mu$-like events, there is both zenith angle (path
length) and energy dependent as expected from Eq. (20) in the
$\nu_\mu \to \nu_\tau$ oscillation scenario.
A fit to the oscillation hypothesis yields $\sin^2 2\theta = 1.01$
and $\Delta m^2 = 0.0032~\rm eV^2$, with a $\chi^2$ of 135 for
152 degrees of freedom.
This corresponds to maximum mixing with very small mass difference.
The various confidence intervals on the oscillation parameters are shown
in Fig.~\ref{figure:oscillation}.

\subsection{Search for $\nu_\tau$ Oscillation in Short Baseline Experiments}

CHORUS and NOMAD are two short baseline neutrino oscillation
experiments at CERN.
The experiments are designed to search for $\nu_\mu \to \nu_\tau$
oscillation through the observation of charged current interactions
$\nu_\tau N \to \tau^- X$.
The neutrino beam also contains a small fractions of $\bar\nu_\mu$,
$\nu_e$, and $\bar\nu_e$ plus a tiny contamination of prompt
$\nu_\tau$, well below the detectable level.
The experiments can therefore also search for $\nu_e \to \nu_\tau$
oscillation.
The search is sensitive to very small mixing angles for large mass
differences, $\Delta m^2 > 1~\rm eV^2$.
This is a cosmologically interesting region: a neutrino with a mass
in this region is a good candidate for the hot dark matter in the
universe.

\begin{figure}
\centering
\centerline{\hbox{\psfig{figure=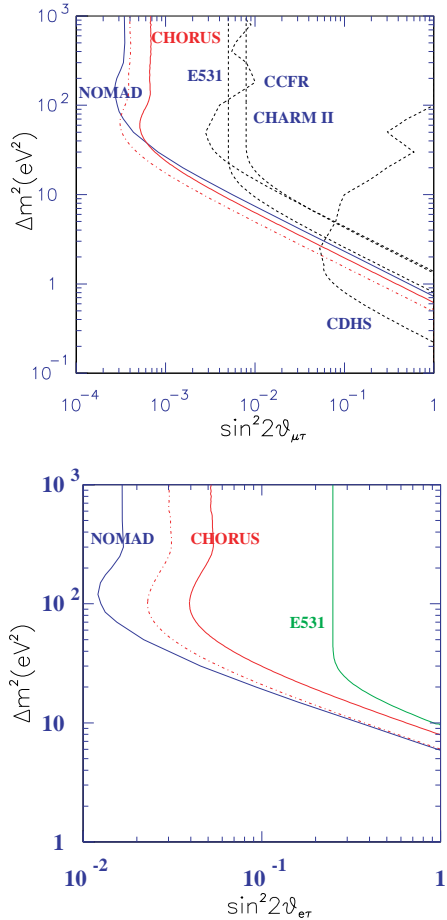, width=7.0cm}}}
\vspace{-0.2in}
\caption{Mixing parameters excluded by various experiments for
         $\nu_\mu \to \nu_\tau$ (top) and $\nu_e \to \nu_\tau$
         (bottom) oscillations.  CHORUS uses the method proposed by
         Junk~\cite{Junk} to combine limits from different decays.
         The limits (dash-dotted) obtained using the Feldman and
         Cousins~\cite{FC} method are also shown for completeness.}      
\label{figure:nu_mu}
\end{figure}

The two experiments deploy very different techniques for the search
of $\tau$ appearance.
CHORUS uses an emulsion target whose excellent spatial resolution
allows a three dimensional visual reconstruction of the decays
$\tau^- \to \mu^-\bar\nu_\mu\nu_\tau$ and $h^-(\pi^0)\nu_\tau$.
NOMAD exploits a purely kinematic technique to identify 
the decays $\tau^- \to e^-\bar\nu_e\nu_\tau$, $h^-(\pi^0)\nu_\tau$,
and $h^-h^+h^-(\pi^0)\nu_\tau$.
Both experiments do not observe any evidence of neutrino oscillation
and exclude the mixing parameters shown in Fig.~\ref{figure:nu_mu}.

\subsection{Search for $\nu_\tau$ Oscillation in Long Baseline Experiments}

The best fit to the atmospheric $\nu_\mu$ deficit is based on the
$\nu_\mu \to \nu_\tau$ oscillation hypothesis.
This is a major challenge to the Standard Model and the result must be
confirmed with an artificial neutrino beam from an accelerator with
beam parameters that can be varied and monitored.
The observed small mass difference (Eqs. (20) and (21)) dictates that
the experiments be located hundreds of kilometers away from the neutrino
source in order to get appreciable $\nu_\mu \to \nu_\tau$ conversion.

The K2K experiment~\cite{K2K} is the first long baseline experiment
conceived for the propose.
A $\nu_\mu$ beam with an average $E_\nu \sim 1~\rm GeV$ from KEK is
directed toward the Super-Kamiokande detector located 250 km away.
There is a miniature Super-Kamiokande detector at the near end at KEK
to monitor the beam.
The experiment searches for a $\nu_\mu$ deficit at the distant
detector because the average beam energy is significantly below
the 3.5~GeV threshold to produce the $\tau$ lepton.
A signal of 27 $\nu_\mu$ events is observed.
The expected signal is $40.3^{+4.7}_{-4.6}$ events under the no
oscillation scenario.
This disfavors the no oscillation scenario at the $2\sigma$ level.
We can expect new result based on a significantly larger data sample
in the near future.

The MINOS experiment~\cite{MINOS} will also look for $\nu_\mu$ disappearance.
The 3-20 GeV beam from Fermilab will be directed toward the
SOUDAN detector at 730 km away.
The detector is based on a magnetic iron tracker.
The higher energy and rate allow MINOS to cover much more than K2K
of the oscillation parameter region favored by the Super-Kamiokande data.
The experiment is expected to start taking data in the year 2003.

Two experiments~\cite{CNGS}, ICANOE and OPERA, have been proposed to
search for $\nu_\tau$ appearance under the CNGS project (CERN
Neutrinos to Gran Sasso).
The $\nu_\mu$ beam with an average energy of 17~GeV will be directed
toward the detector 732~km away.
ICANOE uses a liquid argon Time Projection Chamber (TPC) to reconstruct
the $\tau$ decay with a kinematic technique similar to that of NOMAD.
OPERA reconstructs the $\tau$ decay using an emulsion detector similar
to that used successfully by DONUT to directly observe $\nu_\tau$ for
the first time.
Both experiments are expected to start collecting data in the year 2005.

\section{Search for Lepton Number Violating $\tau$ Decays}

In the Standard Model, there is no symmetry associated with
lepton number and therefore there is no fundamental conservation
law for lepton number; lepton number conservation is an
experimentally observed phenomenon.
Lepton number violation is expected in many extensions of the
Standard Model such as lepto-quarks, SUSY, superstrings, left-right
symmetric models and models which include heavy neutral leptons.
The	predictions typically depend on one or two unknown masses
of new particles and one or two unknown couplings.
Therefore any	null result from a search can only constrain the
parameter space but cannot rule out a particular model.
Nevertheless the search should be pursued vigorously because
of its profound implication on the Standard Model, should a
positive signal be observed.

The $\tau$ lepton is an excellent laboratory for the search
of physics beyond the Standard Model.
Its large mass allows for searches at high momentum transfer with
many decay channels.
The sensitivity may be enhanced because the $\tau$ lepton is a
third generation lepton.
In some models, the coupling may have a mass dependence, e.g.
$\propto m_\tau^5$, resulting in higher sensitivity than searches
using the $\mu$ decay.

\subsection{Implication from Neutrino Observation}

\begin{figure}[htb]
\centering
\centerline{\hbox{\psfig{figure=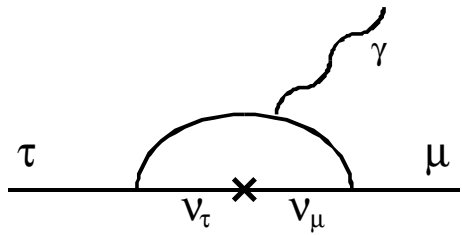, width=6.0cm}}}
\vspace{-0.2in}
\caption{Feymann diagram for the decay $\tau^- \to \mu^-\gamma$
         via neutrino oscillation.}      
\label{figure:mu_g}
\end{figure}

The observed atmospheric $\nu_\mu$ deficit suggests the
$\nu_\mu \to \nu_\tau$ oscillation scenario as discussed in the
previous section.
The effect of neutrino oscillation on the charged lepton number
violating decay has been calculated by Bilenkii and
Pontecorvo~\cite{Pontecorvo} for $\nu_e \to \nu_\mu$ oscillation.
For the $\tau$ decay, the decay rate for $\tau^- \to \mu^-\gamma$
can be calculated using the Feymann diagram shown in
Fig.~\ref{figure:mu_g}:
\begin{eqnarray}
\nonumber
\frac{\Gamma}{\Gamma_{tot}} &\sim& \frac{3\alpha}{32\pi}
     \sin^4(2\theta)\left [\frac{\Delta m^2_\nu}{m^2_W}\right ]^2\\
&=& 5.5 \times 10^{-48}\sin^4(2\theta)(\Delta m^2_\nu)^2
\end{eqnarray}
Using the oscillation parameters extracted by Super-Kamiokande,
this corresponds to a decay rate of $\sim 6 \times 10^{-53}$.
The impact of neutrino oscillation on the neutrinoless decay is
therefore negligible.

There is a new limit on this decay from the CLEO experiment~\cite{mu_g}:
\begin{eqnarray*}
B(\tau^- \to \mu^-\gamma) < 1.1 \times 10^{-6}\ ,
\end{eqnarray*}
at the 90\% CL.
We can expect an order of magnitude improvement of sensitivity
in the decay with the b-factory experiments in the near future.

\subsection{Status of Search for Lepton Number Violating Decays}

The large $\tau$ mass allows the search for lepton number violating
decays in many channels: purely leptonic, radiative decay with a lepton
or proton in the final state, lepton or proton plus pions or kaons.
The experimenters have searched in 49 decay modes and the 90\% CL
upper limits~\cite{PDG} are typically $\sim 10^{-6}$.
The limits on the decays that violate both lepton and baryon numbers
(but conserve baryon minus lepton number) are new since Tau98. 
There are two recent 90\% CL limits from BELLE~\cite{BELLE} with $K^0$:
\begin{eqnarray*}
B(\tau^- \to e^-K^0) < 7.7 \times 10^{-6}\\
B(\tau^- \to \mu^-K^0) < 8.8 \times 10^{-6}
\end{eqnarray*}
These represent significant improvements over the previous
limit~\cite{Hayes} set by MARK~II in 1982.

At this Workshop, Ilakovac~\cite{Ilakovac} presented the predictions
on the branching fractions for lepton number violating decays based
on a model with additional heavy Dirac neutrinos.
Some of the predictions are tantalizingly close to the current
experimental limits, in unit of $10^{-6}$:

\vspace{0.1in}
\(
\begin{array}{lcccc}
                                & & Expt. & & Theory           \\
B(\tau^- \to e^-e^+e^-)         & & < 2.9 & & < 2.7\cdot y^2_{\tau e}\\
B(\tau^- \to e^-\mu^+\mu^-)     & & < 1.8 & & < 1.4\cdot y^2_{\tau e}\\
B(\tau^- \to e^-\pi^0)          & & < 3.7 & & < 2.8\cdot y^2_{\tau e}\\
B(\tau^- \to e^-\rho)           & & < 2.0 & & < 2.7\cdot y^2_{\tau e}\\
B(\tau^- \to e^-\phi)           & & < 6.9 & & < 2.3\cdot y^2_{\tau e}\\
B(\tau^- \to e^-\pi^+\pi^-)     & & < 2.2 & & < 2.7\cdot y^2_{\tau e}\\
B(\tau^- \to e^-K^+K^-)         & & < 6.0 & & < 1.1\cdot y^2_{\tau e}\\
B(\tau^- \to e^-K^0\bar{K^0})   & &       & & < 0.66\cdot z^2_{\tau e}\\
B(\tau^- \to \mu^-K^0\bar{K^0}) & &       & & < 0.13\cdot z^2_{\tau \mu}
\end{array}
\)
\vspace{0.1in}

\noindent
where $y^2_{\tau e}$, $z^2_{\tau e}$, and $z^2_{\tau \mu}$ are
parameters that depend on the masses of the heavy Dirac neutrinos
and their mixings with the ordinary Dirac neutrinos.
It is evident that for two of the related decays, $\tau^- \to e^-\rho$
and $\tau^- \to e^-\pi^+\pi^-$, the experimental limits have already
constrained some of the parameters in the theory.
The last two decays have no experimental limits and are within the
reach of the current b-factory experiments.

\subsection{Future Prospect on Search for Lepton Number Violating Decays}

We can expect another order of magnitude improvement in sensitivity
for the lepton number violating $\tau$ decays in the near future from the b-factory experiments.
This is based on the assumption that the search is not background
limited as was the case with CLEO~II which provides all but four of
the 49 limits.

LHCb~\cite{Bartalini} is an experiment at the Large Hadron Collider
(LHC) at CERN that may have good sensitivity for neutrinoless $\tau$
decays produced in $pp$ collisions at a center-of-mass energy of 14~TeV.
The detector is a single arm spectrometer covering the pseudo-rapidity
region $1.8 < \eta < 4.9$.
The main source of $\tau$ leptons is from the $D_s$ decays,
accounting for 77\% of the total rate.
LHCb plans to collect data at low luminosity, $2 \times 10^{-32}~\rm
cm^{-2}s^{-1}$, in order to reduce occupancy and radiation damage.
This corresponds to the production of $2.5 \times 10^{11}~\tau$'s,
of which $0.7 \times 10^{11}~\tau$'s are in the LHCb acceptance.
This production rate has a large uncertainty because the branching
fraction for $D_s^+ \to \tau^+\nu_\tau$ is poorly known.
The experiment has investigated the sensitivity on the gold-plated
decay, $\tau^- \to \mu^-\mu^+\mu^-$, by requiring three muons with
a detached vertex.
The expected mass resolution is excellent, 5.4 MeV.
However, the potential signal is on a large combinatoric background
of the true muon from $b \to \mu X$ and random tracks.
Under the assumption of no observed signal, the experiment can set a 
90\% CL limit of $1.8 \times 10^{-7}$ after one year of data collection.
This is not very competitive in view of the fact that the b-factory
experiments will reach this sensitivity in the near future and LHCb
is not scheduled to commence data collection until the year 2005.
The estimate of the sensitivity is preliminary  and the search
technique is currently being refined.
Possible modification to the trigger to enhance the sensitivity is
being considered.
The sensitivity to other neutrinoless $\tau$ decays with pions
and kaons will be investigated.

I would like to repeat the suggestion I made at Tau98:
Both CDF and D0 may have good sensitivity to neutrinoless $\tau$ decays
and I would like to urge the experimenters to perform the search!

\section{Future Prospect}

In this section, I will discuss the future prospects of the physics
of the $\tau$ lepton.
The future prospect for the $\tau$ neutrino has already been discussed
in the section on neutrino oscillation.

In the next few years, we expect most of the results will come from the
b-factory experiments.
Both BABAR at SLAC and BELLE at KEK have collected a data sample
comparable to the CLEO~II experiment at Cornell.
The CLEO~II experiment collected $13.5~\rm fb^{-1}$ of data, corresponding
to $1.23 \times 10^{7}~\tau^+\tau^-$ produced events.
We expect most of the analyses performed by CLEO~II will be repeated
with smaller statistical errors.
It will require care and hard work to reduce the systematic errors in
the systematic limited analyses.
In the following, I will discuss a few analyses that we can expect
to see significant progress in the near future:

\begin{itemize}
\item
Michel Parameters: We can expect significant improvement in the
measurement of the Michel parameters with the much larger data samples.
In particular, the measurement of the low-energy parameter $\eta$
will be greatly improved.
As the name implied, the biggest sensitivity to this parameter is
at low lepton energy and the parameter can only be measured in the
muon decay channel because of the helicity flipping factor
$m_l/m_\tau$ for the daughter charged lepton in Eq. (12)
as discussed in Section 3.
The CLEO experiment can only identify muons with momentum above 1 GeV.
The muon detector of BABAR, for example, can identify muons with
momentum as low as 0.6 GeV, greatly increasing the sensitivity to $\eta$.
\item
Cabbibo-Suppressed Decays:
The single most important difference between the CLEO~II and b-factory
detectors is the addition of an excellent kaon identification system.
This will allow precision measurements of major kaon channels and
observation of new rare kaon modes, reminiscence of the advance in
the study of decays with $\pi^0$ and $\eta$'s provided by the CsI
calorimeter of the CLEO~II detector.
\item
Second Class Current:
Observation of the second class current decay $\tau^- \to \pi^-\eta\nu_\tau$
is one of the major goals of $\tau$ physics of the b-factory experiments.
The branching fraction of the decay is expected to be
$(1.2-1.5)\times 10^{-5}$~\cite{Pich}.
The upper limit on the branching fraction as extracted by the CLEO~II
experiment~\cite{Keta} is
$B(\tau^- \to \pi^-\eta \nu_\tau) < 1.4 \times 10^{-4}$ at 95\% CL.
In the analysis, the major background with an $\eta$ signal is from
the hadronic events ($e^+e^- \to q\bar q$) and the decays
$\tau^- \to \pi^-\pi^0\eta\nu_\tau$ and $K^-\eta\nu_\tau$.
The latter decay has the smallest contribution and can be easily
eliminated with the excellent kaon identification capacity.
The other two backgrounds can be suppressed with a much tighter photon veto.
This selection criterion was not imposed in the CLEO~II
analysis because the analysis was designed to study the decay
$\tau^- \to K^-\eta\nu_\tau$ which had a different background contamination.
The second class current decay can therefore be observed in the near future.
\item
$\nu_\tau$ Mass: The atmospheric $\nu_\mu$ deficit observed by
Super-Kamiokande favors the solution of $\nu_\mu \to \nu_\tau$
oscillation with $\Delta m^2 = 0.0032~\rm eV^2$.
Since the upper limit on the $\nu_\mu$ mass is 0.19~MeV at
the 90\% CL~\cite{PDG}, this implies that the $\nu_\tau$ mass
must also be less than 0.19~MeV, beyond the sensitivity of
any b-factory experiment.
However, as discussed in Section 9, the confirmation of
the $\nu_\mu \to \nu_\tau$ oscillation hypothesis with the
detection of $\nu_\tau$ appearance will have to wait until
the year 2005.
The experimenters should therefore continue to measure the
$\nu_\tau$ mass with the much larger data sample.
With the excellent kaon identification, the use of
kaon decay modes such as $\tau^- \to \pi^-K^+K^-\nu_\tau$
and $K^-K^+K^-\nu_\tau$ could greatly enhance the sensitivity.
With luck, a sensitivity to $\nu_\tau$ mass as low as 10~MeV
is quite possible.
\end{itemize}
 
\section{Conclusion}

A large number of results were presented at Tau2000.
The Standard Model is being tested both with the $\tau$ lepton as a
decay product and with the decay products of the $\tau$ lepton.
This includes the test of lepton universality, measurement of the
Lorentz structure, and search for lepton number violating decays.
There is no hint of physics beyond the Standard Model.
The $\tau$ lepton is also used as a laboratory to measure
the strong coupling constant, strange quark mass, and the
hadronic vacuum polarization, in addition to the test of QCD.

The neutral partner of the $\tau$ lepton, $\nu_\tau$, on the
other hand, points to physics beyond the Standard Model via
the observation of $\nu_\mu \to \nu_\tau$ oscillation.
This implies the violation of lepton number conservation, a
fundamental assumption of the Standard Model.
This challenge to the Standard Model needs to be confirmed with
the direct observation of $\nu_\tau$ appearance in an $\nu_\mu$
beam, an observation that is now proven to be possible by the
direct observation of $\nu_\tau$ by DONUT.

This is the ten-year anniversary of the Workshop dedicated to examine
the $\tau$ lepton and its neutrino.
There has been tremendous progress in the field.
The first Workshop marked the transition from DORIS/PEP/PETRA to CESR/LEP.
This Workshop marks the transition to the b-factory and
neutrino oscillation experiments.
LEP is a laboratory where the major decay modes of the $\tau$ lepton
can be studied with high efficiency and hence low systematic error.
CLEO~II excels in decays with $\pi^0$ and $\eta$ and dominates the
search for lepton number violating decays via its large data sample.
The b-factory experiments will collect a significantly larger data
sample with excellent kaon identification, opening a new window
of opportunity to challenge the Standard Model.

\vspace{0.1in}
\noindent
{\bf Acknowledgements}

\vspace{0.1in}
This work was supported in part by the U.S.~Department of Energy.
The author wishes to thank all the speakers at Tau2000 that provided
the advance information on their talks to allow him to put
together the conference summary talk.
The author also wishes to acknowledge the many useful discussions
with R.D.~Kass, M. Davier, L. Di Lella, S. Eidelman, A. Ilakovac,
K. Maltman, and M.L. Perl.
The author wishes to congratulate M. Roney and R. Sobie for a
successful conference.
Special thanks go to K. Graham for the ride from the banquet
to the hotel so that the author could prepare for the  conference summary
talk of the following day.


\begin{thebibliography}{9}
\bibitem{Marciano} W.J. Marciano and A. Sirlin, Phys. Rev. Lett. 61 (1988) 1815.

\bibitem{Robertson} S.H. Robertson, these proceedings.

\bibitem{Bella} G. Bella, these proceedings.

\bibitem{PDG} D.E. Groom \etal~(Particle Data Group), Eur. Phys. J. C 15 (2000) 1.

\bibitem{Reid} D.W. Reid, these proceedings.

\bibitem{EWWG} See the results of summer 2000 of the LEP Electroweak Working Group
               at http:// www.cern.ch/LEPEWWG.

\bibitem{Michel} L. Michel, Proc. Phys. Soc. A 63 (1950) 514, 1371;
                 C. Bouchiat and L. Michel, Phys. Rev. 106 (1957) 170;
                 T. Kinoshita and A. Sirlin, Phys. Rev. 107 (1957) 593, 108 (1957) 844;
                 W. Fetscher, H.-J. Gerber, ad K.F. Johnson, Phys. Lett. B 173 (1986) 102.

\bibitem{Boyko} I. Boyko, these proceedings.

\bibitem{Weinberg} S.~Weinberg, Phys.~Rev. 112 (1958) 1375.

\bibitem{Feynman} R.~P.~Feynman and M.~Gell-Mann, Phys. Rev. 109 (1958) 193.

\bibitem{Eidelman98} S.I. Eidelman and V.N. Ivanchenko, Nucl. Phys. B (Proc. Suppl.)
                     76 (1999) 319.

\bibitem{PDG98} There is no difference between 1998 and 2000~\cite{PDG}
                Particle Data Group world average value of $B_{\pi\pi}$.

\bibitem{Chen} G.M. Chen, these proceedings.

\bibitem{Eidelman} S.I. Eidelman, these proceedings.

\bibitem{4pi} K.W. Edwards \etal, Phys. Rev. D 61 (2000) 072003.

\bibitem{2pi} S.~Anderson \etal, Phys.~Rev. D 61 (2000) 112002.

\bibitem{6pi} A. Anastassov \etal, CLNS 00/1687, submitted to Phys. Rev. Lett.

\bibitem{GaoAndLi} J. Gao and B.A. Li, hep-ph/0004097.

\bibitem{3pieta} T.~Bergfeld \etal, Phys. Rev. Lett. 79 (1997) 2406.

\bibitem{Pais}   A.~Pais, Ann. Phys. 9 (1960) 548.

\bibitem{Eidelman97} S.I.~Eidelman and V.N.~Ivanchenco, Nucl. Phys. B (Proc. Suppl.) 55C
                     (1997) 181.

\bibitem{Davier} R. Alemany, M. Davier, and A. H\"{o}cker, Eur. Phys. J. C 2 (1998) 123.

\bibitem{g-2} B. Lee Roberts, Z. Phys. C 56 (Proc. Suppl.) (1992) 101;
              see also http://www.phy.bnl.gov/ g2muon/home.html.

\bibitem{ms} M. Davier \etal, these proceedings.

\bibitem{Maltman} K. Maltman and J. Kambor, these proceedings.

\bibitem{Lubicz} V. Lubicz, hep-lat/0012003.

\bibitem{Samuel} M.A. Samuel, G.W. Li, and R. Mendel, Phys. Rev. Lett. 67 (1991) 668.

\bibitem{Bernabeu} J. Bernabeu, G.A. Gonzalez-Sprinberg, M.~Tung,
                   and J. Vidal, Nucl. Phys. B 436 (1995) 474.

\bibitem{ARGUS} H. Albrecht \etal, Phys. Lett. B 485 (2000) 37.

\bibitem{L3} M. Acciarri \etal, Phys. Lett. B 434 (1998) 169.

\bibitem{DONUT} B. Baller, these proceedings.

\bibitem{SuperK} R. Svoboda, these proceedings.

\bibitem{Shibuya} H. Shibuya, these proceedings.

\bibitem{CHOOZ} M. Apollonio \etal, Phys. Lett. B 466 (1999) 415.

\bibitem{Palo_Verde} F. Boehm \etal, Phys. Rev. Lett. 84 (2000) 3764.

\bibitem{Junk} T. Junk, Nucl. Instr. and Meth. A 434 (1999) 435.

\bibitem{FC} G.J. Feldman and R.D. Cousins, Phys. Rev. D 57 (1998) 3873.

\bibitem{K2K} S. Boyd, these proceedings.

\bibitem{MINOS} A. Weber, these proceedings.

\bibitem{CNGS} L. Di Lella, these proceedings.

\bibitem{Pontecorvo} S.M. Bilenky and B. Pontecorvo, Phys. Lett. B 61 (1976) 248.

\bibitem{mu_g} S. Ahmad \etal, Phys. Rev. D 61 (2000) 071101;
               R.D. Kass, these proceedings.

\bibitem{BELLE} A. Abashian \etal, BELLE-CONF-0016, submitted to XXXth Int.
                Conf. High Energy Phys., Osaka, Japan, 2000.

\bibitem{Hayes} K.G. Hayes, Phys. Rev. D 25 (1982) 2869.

\bibitem{Ilakovac} A. Ilakovac, these proceedings.

\bibitem{Bartalini} P. Bartalini, these proceedings.

\bibitem{Pich} A. Pich, Phys. Lett. B 196 (1987) 561;
               S. Tisserant and T.N. Truong, Phys. Lett. B 115 (1982) 264;
               H. Neufeld and H. Rupertsberger, Z.~Phys. C 68 (1995) 91.

\bibitem{Keta} J.~Bartelt \etal, Phys.~Rev.~Lett. 76 (1996) 4119.

\end{thebibliography}
\end{document}